# Coherent Hopping Transport and Giant Negative Magnetoresistance in Epitaxial CsSnBr$_3$

Liangji Zhang,[1] Isaac King,[2] Kostyantyn Nasyedkin,[1,3] Pei Chen,[2] Brian Skinner,[4] Richard R. Lunt,[2,*] and Johannes Pollanen[1,*]

[1]Department of Physics and Astronomy, Michigan State University, East Lansing, MI 48824, USA
[2]Department of Chemical Engineering and Materials Science, East Lansing, MI 48824, USA
[3]Neutron Scattering Division, Oak Ridge National Laboratory, Oak Ridge, TN 37831, USA
[4]Department of Physics, Ohio State University, Columbus, Ohio 43210, USA

**Abstract:** Single-crystal inorganic halide perovskites are attracting interest for quantum device applications. Here we present low-temperature quantum magnetotransport measurements on thin film devices of epitaxial single-crystal CsSnBr$_3$, which exhibit two-dimensional Mott variable range hopping (VRH) and giant negative magnetoresistance. These findings are described by a model for quantum interference between different directed hopping paths and we extract the temperature-dependent hopping length of charge carriers, their localization length, and a lower bound for their phase coherence length of ~100 nm at low temperatures. These observations demonstrate that epitaxial halide perovskite devices are emerging as a material class for low-dimensional quantum coherent transport devices.

Halide perovskites have emerged over the past decade as a fundamentally intriguing class of semiconductors with a wide variety of potential applications. These materials have shown remarkable optoelectronic properties including high photoluminescence quantum yields[1], high optical absorption[2], long carrier diffusion length and widely tunable bandgaps[3]. These optical and electronic properties make halide perovskites excellent for electronic devices such as light emitting diodes [4], semiconductor lasers [5], and solar cells[6]. To overcome the limitations of grain boundaries and ionic defects common in conventional solution-processed polycrystalline halide perovskite, single-crystal epitaxial growth of halide perovskite has been introduced[7–11]. These new growth techniques also enable the study of delicate quantum properties of charge carriers in halide perovskites in quantum devices and 2D quantum wells[7]. Previously, we have reported on the observation of phase coherent transport of charge carriers in the presence of spin-orbit coupling manifesting in weak anti-localization in epitaxial CsSnI$_3$[12]. Phase coherent quantum interference effects in the form of weak localization have also been recently reported in quasi-epitaxial CsPbBr$_3$[13]. Additionally, hybrid organic-inorganic halide perovskites of single crystal MAPbI$_3$ and MAPbBr$_3$ have been shown to exhibit a variety of low-temperature and high-magnetic field transport phenomena under illumination[14]. In this present work we show that coherent transport effects can be realized in another high-quality halide perovskite epitaxial material, namely thin films of CsSnBr$_3$, at low temperature and high magnetic field. These effects result from the interference of coherent charge carrier hopping trajectories and manifest in a large device magnetoresistance. In addition to enhancing the fundamental understanding of low-dimensional phenomena in halide perovskites, low-temperature magnetoresistance measurements such as those described in this manuscript also provide key insights into the possible transport regimes that can realized in this class of materials. Looking ahead, it will be compelling to investigate which material parameters (e.g. doping, crystal structure, strain, etc.) are involved in transitioning between different transport regimes given that a variety of coherent effects have now been observed[12–14]. This knowledge will impact the material design and development of future quantum electronic devices based on epitaxial halide perovskites.

Two-dimensional (2D) Mott variable range hopping (VRH) is a mechanism for transport in a disordered 2D electron system in which the wave functions of charge carriers tend to localize in the vicinity of lattice defects. At sufficiently low temperature, the electric current in these systems is transported by phonon-assisted tunneling from one localized state to the other, with the dominant hopping processes taking place between distant rather than nearest-neighboring sites[15,16]. The hopping rate between localization sites leads to Mott's law for conductivity, which in 2D systems is given by $\ln[R(T)] \sim (1/T)^{1/3}$. Many studies have reported that systems in the VRH regime can also show large negative magnetoresistance, including gallium arsenide (GaAs) field-effect transistors[17], GaAs/AlGaAs heterostructures[18], fluorinated graphene[19], Ge films[20] and In$_2$O$_{3-x}$ films[21]. In order to explain the negative magnetoresistance, Nguyen, Spivak and Shkovskii developed a model based on the interference of coherent tunneling trajectories of charge carriers, i.e. the so-called NSS model[22,23]. They found that in the VRH regime, the quantum interference between distinct forward hopping paths leads to a suppression of the hopping rate between distant sites. Negative magnetoresistance results from the alteration of the relative phase between different hopping paths by the magnetic field, analogous to the effect of weak localization (WL) in the quantum diffusive transport regime. In this paper, we find evidence of this effect in epitaxial CsSnBr$_3$ thin films. The temperature dependent resistance of our single-crystal epitaxial CsSnBr$_3$ system matches well with the 2D Mott VRH and the giant negative magnetoresistance we observe fits the existing theoretical NSS model predictions. Furthermore, this theoretical model allows us to estimate the various length scales associated with the quantum coherent hopping of charges in this system and place a lower bound on the phase coherence length.

We begin by describing the physics underlying the NSS model and its resemblance to the theory of WL in quantum diffusive transport. In Fig. 1 we present a qualitative picture of these two modes of quantum coherent transport. The various relevant length scales, including the temperature-dependent typical hopping length $r(T)$, the charge carrier localization length at zero magnetic field $\xi$, and the phase coherence length $L_\Phi$ are shown in Fig. 1 along with their sizes relative to each other.

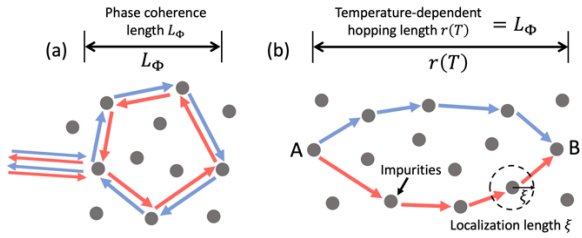

FIG. 1. Qualitative picture of two modes of quantum coherent transport in 2D systems. The grey dots represent impurities (sites of elastic scattering). Three different length scales, the phase coherence length $L_\Phi$, the temperature-dependent hopping length $r(T)$, and the localization length $\xi$, are indicated. (a) Model of WL in a conductor. Blue (red) arrows show clockwise (counter-clockwise (time-reversed)) diffusive trajectories. (b) VRH and the NSS model in an insulator. A charge carrier tunnels from site A to an energetically favorable site B. The red path and the blue path represent two possible tunneling trajectories, which contribute to the overall hopping amplitude from A to B.

In the quantum diffusive transport regime, which arises at a higher charge carrier density relative to the VRH-regime, the coherence among multiple elastic scattering paths of a single electron can lead to an enhancement of backscattering, as shown in Fig. 1 (a). An externally applied magnetic field introduces relative phase shifts between different scattering paths and thus leads to negative magnetoresistance, which is the hallmark of WL [24]. In analogy to WL, coherent transport also appears at significantly lower charge carrier density in the variable range hopping regime and can be affected by a magnetic field. To understand the magnetoresistance in the hopping regime, the NSS model considers the overall hopping amplitude between two localized sites as the sum of amplitudes of different tunneling trajectories, each of which involves an electron or hole passing virtually through multiple localized impurity states, as depicted in Fig. 1 (b). When an external magnetic field is applied, a phase shift is introduced that can coherently enhance the overall hopping amplitude and lead to a large negative magnetoresistance [25,26]. In this work, we present magnetotransport measurements on epitaxial CsSnBr$_3$ devices demonstrating VRH coherent transport that is consistent with the theoretical description provided by the NSS model.

For these measurements a 30 nm thick epitaxial film of CsSnBr$_3$ was deposited stoichiometrically on a cleaved [100] surface of a sodium chloride (NaCl) single crystal substrate. The growth of the epitaxial layer was performed at pressures less than $3\times10^{-6}$ torr and a temperature of 23°C. The schematic structure of the epilayer and in-situ real time reflection high energy electron diffraction patterns of the grown epilayer are shown in Figure 2. We note that streaks for the perovskite layer emerge between the substrate streaks due to the transition from the face-centered cubic structure to a primitive one. After growth, 2.0 mm × 0.5 mm gold pads (100 nm thick) were electron-beam evaporated onto the CsSnBr$_3$ epilayer to enable standard lock-in based low-frequency (10 Hz) ac electrical transport measurements (Fig. 2). After attaching measurement leads, devices were loaded into a hermetically indium o-ring sealed copper sample container and thermally anchored to the mixing chamber of a dilution refrigerator for the low temperature (~10 mK) and high magnetic field (up to $B = \pm13.5$ ) transport measurements. Due to the air-sensitive nature of the samples, devices were kept in an entirely dry, oxygen-free environment during fabrication, wire-up, cool down and measurement (see Supporting Information section S1).

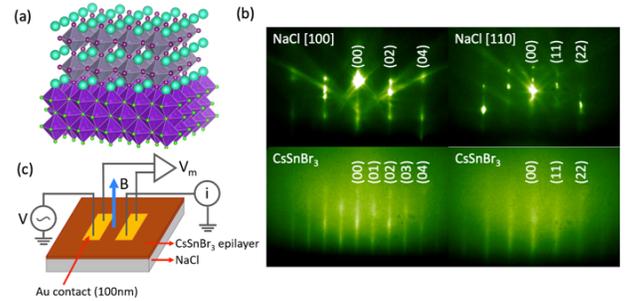

FIG. 2 (a) Crystal structure of epitaxial cesium tin bromide (CsSnBr$_3$) on NaCl (cyan is Cs, gray is Sn, dark purple is Br, green is Na, purple is Cl). (b) Reflective high-energy electron diffraction (RHEED) pattern of the NaCl substrate (top panels) and CsSnBr$_3$ thin film showing crystalline streaks in both the NaCl and epitaxially grown CsSnBr$_3$ that vary as expected with rotation. Note the additional (01) streaks emerge for CsSnBr$_3$ due to the primitive cell versus the face-centered cubic cell of the substrate. (c) Schematic of the transport measurement setup. These experiments were performed between evaporated gold contacts on 30 nm thick epitaxial CsSnBr$_3$ thin film devices. The device conductivity was calculated from the measured value of the voltage $V_m$ and the measured current I, which were obtained using standard low-frequency AC lock-in. A magnetic field, B, perpendicular to the plane of the CsSnBr$_3$ epilayer enabled measurement of the magnetotransport at low temperature.

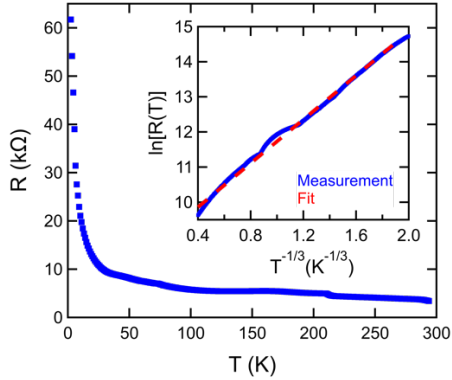

FIG. 3. Temperature dependent resistance of CsSnBr$_3$ epitaxial thin film at B = 0. Inset: Logarithm of the resistance as a function of $T^{-1/3}$. The red line is a fit to the data based on the temperature dependence of 2D Mott VRH transport.

In Fig. 3 we show the resistance of a CsSnBr$_3$ epitaxial thin film device versus temperature as it is cooled down from room temperature to 2.4 K in the absence of a magnetic field. We observe that the device resistance increases as the temperature decreases and reaches many tens of kΩ at the lowest temperatures. This change is consistent with that of a lightly-doped semiconductor. While previous work has reported ohmic contact between CsSnBr$_3$ and Au at room temperature[7,27], our two-terminal resistance measurements cannot preclude a contribution from a contact resistance between the evaporated gold and the CsSnBr$_3$ epilayer, however we emphasize that such a contact resistance would not change the conclusions we draw regarding coherent hopping transport described below. We also note that recent theoretical work has indicated the potential role of electrode induced impurities in CsSnBr$_3$[28], which could be investigated with future 4-terminal devices. Finally we note that all measurements reported here were performed in a regime of linear-response, which was ensured by performing device $IV$-characteristics at the lowest temperatures.

We note that a small kink in the resistance appears at ~215 K in all of the samples we have measured (see Fig. 3 and Supporting Information section 2). It is likely associated with a structural phase transition in the epilayer. In fact, structural phases transitions have been reported previously in bulk CsSnBr$_3$[29]. Moreover, our *in-situ* RHEED diffraction measurements on rough CsSnBr$_3$ shows an increase in the c/a lattice ratio of $7 \pm 1$ % as the sample is cooled from room temperature to 83 K, confirming the likely presence of a cubic to tetragonal phase transition (see Supporting Information section 2).

In the inset of Fig. 3, the low-temperature data show the characteristic dependence associated with 2D Mott VRH transport and are well-fit to the form of $\ln[R(T)] \sim (T_0/T)^{1/3}$[30]. The characteristic temperature $T_0 =$ $32\ K$ is obtained from the slope of the fitted line. The observation of Mott VRH from CsSnBr$_3$ epitaxial thin film device indicates that at low temperatures, the tunneling distance of strongly localized charge carriers in CsSnBr$_3$ is much larger than the distance between impurities. The electrons undergo phonon assisted tunneling through intermediate localized states, as depicted in Fig. 1 (b). As the temperature is increased we observe the expected crossover from VRH to thermally activated transport, while below roughly 125 mK we observe a saturation of the device resistance that is likely associated with thermal decoupling of the device (see Supporting Information section S3).

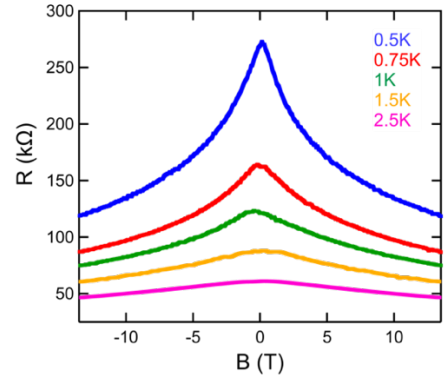

FIG. 4. Magnetoresistance of a CsSnBr$_3$ epitaxial thin film at different temperatures. The applied magnetic field is normal to the CsSnBr$_3$ film. The peak of the magnetoresistance weakens as the temperature increases, indicating that the length scale associated with phase-coherent hopping is reduced. We note that the data presented here is from a single device but has been reproduced with additional samples grown separately (see Supporting Information section S4).

The coherent interference between hopping transport trajectories can be revealed by performing device resistance measurements under the application of a large magnetic field. In Fig. 4 we present results of the low-temperature magnetoresistance measurements on epitaxially grown CsSnBr$_3$. The measurements were performed at different temperatures with an external magnetic field applied normal to the plane of the epitaxial film. Unlike the negative magnetoresistance often observed in the WL regime, in which the magnetoresistance only varies by a few percent, we observe a giant negative magnetoresistance with no sign of saturation over the full field range (±13.5 T) at $T = 0.5\ K$ with the ratio of $R(0)/R(B)$ significantly exceeding unity. This is consistent with results shown in reports for GaAs/AlGaAs heterojunctions and In$_2$O$_{3-x}$ films in the VRH regime[18,21], and consistent with the predictions of the NSS model. We also observe a weakening of the negative magnetoresistance peak as the measurement temperature increases, which could indicate that the length scale associated with phase-coherent hopping is reduced. This could result from increased inelastic scattering from phonons or charge carriers. Alternatively this weakening could potentially be associated with a contribution from positive (orbital) magnetoresistance as the temperature is increased.

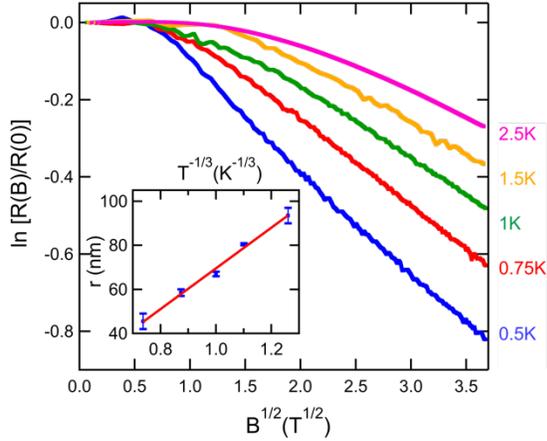

FIG. 5. Logarithm of the normalized magnetoresistance $\ln[R(B)/R(0)]$ as a function of the square root of the magnetic field $(B^{1/2})$ at different temperatures as indicated. Inset: temperature-dependent hopping length $r(T)$ as a function of $T^{-1/3}$.

The NSS-model also provides a framework for understanding the functional dependence of the magnetoresistance as the temperature is increased [26], which is given by,

$$\ln\left[\frac{R(B)}{R(0)}\right] = -\gamma r(T)\left(\frac{e}{\hbar}\right)^{\frac{1}{2}} B^{\frac{1}{2}} \quad (1)$$

where $\gamma$ is a numerical coefficient which is estimated to be $0.1/\sqrt{2}$ and $r(T)$ is the temperature-dependent hopping length. In particular, Eq. 1 predicts a characteristic square-root dependence of $\ln R(B)$, which serves as a fingerprint of the underlying NSS physics. We have plotted the magnetoresistance data in the form of natural logarithm of normalized magnetoresistance $\ln[R(B)/R(0)]$ as a function of $B^{1/2}$ in Fig. 5, which clearly demonstrates the expected functional dependence. The temperature dependent hopping length $r(T)$ can be obtained from the slope of the linear regime in Fig. 5 (see Eq. 1), which yields $r(T) \cong 94$ nm, 81 nm, 67 nm, 59 nm and 46 nm at 0.5 K, 0.75 K, 1 K, 1.5 K and 2.5 K, respectively. In the low temperature regime of 2D Mott VRH, $r(T)$ should have the following temperature dependence $r(T) \approx \xi(T_0/T)^{1/3}$ [15,16]. In the inset of Fig. 5, we plot $r(T)$ as a function of $T^{-1/3}$, which shows good agreement with this expected behavior for 2D Mott VRH and is independently consistent with our temperature dependent resistance measurements at zero magnetic field. We also note that the corresponding localization length for charge carriers $\xi$ can be determined from the slope of $r(T)$, from which we find $\xi \approx 30$ nm.

Our measurements and analysis also allow us to place a lower bound on the phase coherence length for charge carriers in CsSnBr$_3$. In particular, the NSS model is applicable when the phase coherence length L$_\Phi$ is longer than $r(T)$, as shown in Fig. 1 (b). In this limit the hopping paths retain phase coherence. Eq. (1) also requires that the magnetic length $L_B = (\hbar/eB)^{1/2}$ be shorter than the hopping length $r(T)$, so that there are multiple quanta of magnetic flux through a typical closed-loop hopping path, and the magnetic field strongly alters the phase of the charge carrier trajectories. Our observation in aggregate indicate good agreement with the NSS-model suggesting that both conditions are fulfilled, and the hierarchy of length scales $L_B \leq r(T) \leq L_\Phi$ is achieved. In Fig. 5 at $T = 0.5\ K$, we observe linear dependence of $\ln[R(B)/R(0)]$ on $B^{1/2}$ for all values of $B^{1/2} \geq 0.6\ T^{1/2}$, with a weaker dependence at smaller fields. This high-field regime corresponds to $L_B \leq 40\ nm$, and the condition $L_B \leq r(T = 0.5\ K)$ is satisfied. While we cannot directly measure the phase coherence length L$_\Phi$ in the VRH regime, the requirement that L$_\Phi$ is greater than $r(T = 0.5\ K)$ implies a phase coherence length significantly larger than at least ~ 100 nm. This is consistent with the notably high phase coherence length measured in epitaxial CsSnI$_3$ [12].

In summary, we have observed giant negative magnetoresistance in single-crystalline epitaxial thin film CsSnBr$_3$. The devices exhibit a strongly localized 2D Mott VRH of charge carriers. The negative magnetoresistance can be understood within the context of the NSS model, which describes the interference of coherent hopping between localized states. Based on this model we analyze our data to determine the relevant transport length scales including the low temperature hopping length and the localization length. Additionally, we are able to place a lower bound on the phase coherence length of ~100 nm for charge carriers in this epitaxial thin film halide perovskite. To our knowledge these results are the first to demonstrate phase coherent hopping transport and the resulting large negative magnetoresistance in halide perovskites. They add to the growing body of evidence demonstrating that epitaxial halide perovskite thin film devices are emerging as an exciting new class of low-dimensional quantum electronic materials. As the quality of epitaxial halide perovskite materials continues to improve via advances in growth, doping and strain engineering one can expect the emergence of ever more subtle and exotic electronic states of matter in these materials at low-temperatures. For example, future low-dimensional electron systems based on halide perovskites such as CsSnBr$_3$ could one day provide a novel platform for investigating the collective states in the fractional quantum Hall regime. Additionally, leveraging the advances of halide perovskite epitaxy opens the door for developing high-quality devices exhibiting symmetry-broken interfacial states with unique physical properties (e.g. ferroelectricity, coherent transport, spin-orbit coupling) not realizable in bulk materials.

We thank J.I.A. Li, B.I. Shklovskii and E.A. Henriksen for fruitful discussions. This work was supported by the National Science Foundation via grant no. DMR-1807573. J.P. also acknowledges the valuable support of the Cowen Family Endowment at MSU. B.S. was partially supported by the Center for Emergent Materials, an NSF-funded MRSEC, under Grant No. DMR-2011876.

## ASSOCIATED CONTENT

### Supporting information

$CsSnBr_3$ device handling details; evidence for a structural phase transition at low temperature; temperature dependent device transport outside the VRH regime; data for additional device

* rlunt@msu.edu
* pollanen@msu.edu

# Supporting Information

Coherent Hopping Transport and Giant Negative Magnetoresistance in Epitaxial CsSnBr$_3$

Liangji Zhang,[1] Isaac King,[2] Kostyantyn Nasyedkin,[1,3] Pei Chen,[2] Brian Skinner,[4]

Richard R. Lunt,[2,*] and Johannes Pollanen[1,*]

[1]Department of Physics and Astronomy, Michigan State University, East Lansing, M 48824, USA

[2]Department of Chemical Engineering and Materials Science, East Lansing, MI 48824, USA

[3]Neutron Scattering Division, Oak Ridge National Laboratory, Oak Ridge, TN 37831, USA

[4]Department of Physics, Ohio State University, Columbus, Ohio 43210, USA

* rlunt@msu.edu
* pollanen@msu.edu

## SI 1: CsSnBr$_3$ device handling details:

After growth, the CsSnBr$_3$ samples were sealed in a dry nitrogen environment inside of a KF vacuum clamp assembly, which was transported to the low-temperature quantum measurement lab and placed in dry nitrogen filled enclosure where the sample was wired onto an 18-pin chip holder and mounted inside a hermetically indium o-ring sealed copper sample mount (see Fig. S6 below), while still in dry nitrogen. This sample mount was attached to the cold-finger of the cryogen free-dilution refrigerator equipped with a 14T dry superconducting magnet.

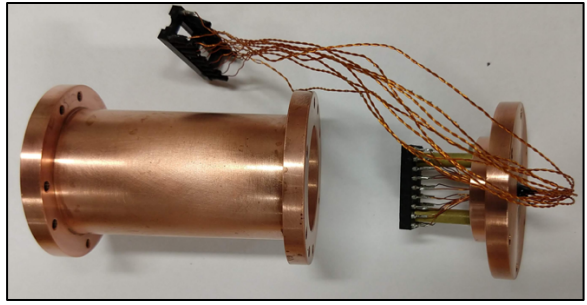

**Figure S1 | Hermetically sealed sample mount.** Devices are sealed in a sample cell in a dry nitrogen environment.

## SI 2: Evidence for a structural phase transition at low temperature:

As shown in Fig. S2, we have observed a small kink-like feature in the resistance of all devices at a temperature of approximately 215 K. It is likely associated with a structural phase transition in the CsSnBr$_3$ epilayer. In fact, structural phase transitions have been previously observed in bulk CsSnBr$_3$ (see Ref. [29] of the manuscript) albeit at slightly higher temperature. It is possible that the epitaxial registry with the substrate could stabilize the transition to lower temperature. Further corroborating the existence of a structural phase transition are our in-situ RHEED data on rough CsSnBr$_3$ shown in Fig. S3. We find an increase in the c/a lattice ratio by $7 \pm 1\%$ as the sample is cooled from room temperature to 83 K (liquid nitrogen cooling), indicative of a cubic to tetragonal phase transition.

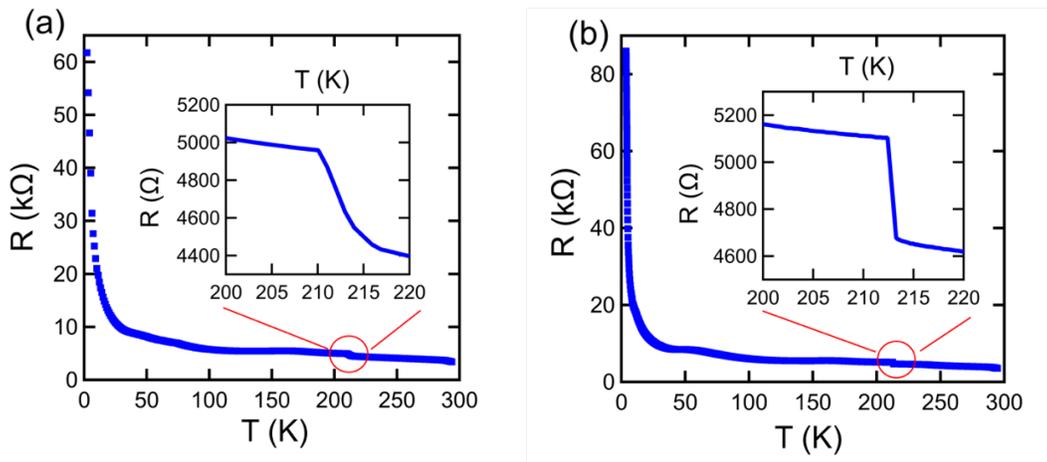

**Figure S2 | Temperature dependent resistance of additional CsSnBr$_3$ device at B = 0.** (a) Original CsSnBr$_3$ device presented in the main manuscript. (b) additional CsSnBr$_3$ device for comparison. Insets: Data zoomed in near 215 K, highlighting the observed kink in $R(T)$, which is indicative of a structural phase transition upon cooling.

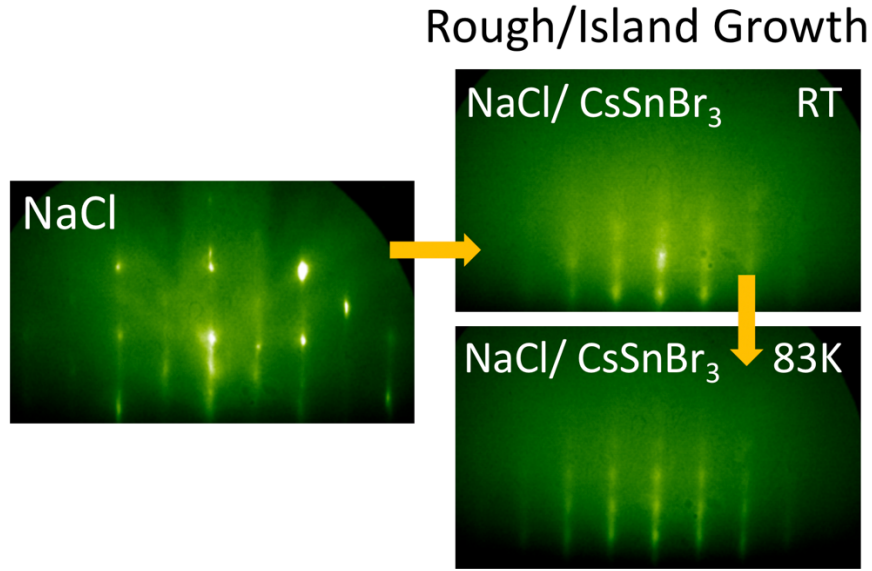

**Figure S3 | Evidence for structural phase transition via in-situ diffraction.** *In-situ* RHEED patterns of NaCl (left) and with epitaxial CsSnBr$_3$ grown under conditions to be rough to enable c-lattice parameter information to be discernable, cooled from room temperature (top, right) to just above LN$_2$ temperature (bottom, right). When the film is cooled to 83 K, we find an increase in the c/a ratio of ~7 +/- 1% based on the change in the spot spacing in the *dz (vertical)/dx (horizontal)* directions. This is indicative of a cubic to tetragonal phase transformation that occurs between room temperature and 83 K, which is consistent with the kink in the temperature dependent device resistance measurements that suggest that this transition occurs near 215 K.

**SI 3: Temperature dependent device transport outside of the variable range hopping regime:**

The regime of variable range hopping (VRH) will only appear at a sufficiently low temperature. As the temperature increases thermally activated transport will onset and dominate hopping transport produced by tunneling between localized sites. Thermally activated transport is characterized by a dependence $\ln[R(T)] \propto 1/T$. In Fig. S4(a) we present measurements of the device resistance at higher temperature showing this expected $1/T$ dependence for our epitaxial CsSnBr$_3$ device.

Additionally we observe a saturation of the device resistance roughly 125 mK as shown in Fig. S4(b). This deviation is likely associated with a thermal decoupling of the device from the mixing chamber as the primary source of electronic heat exchange between the sample and the cryostat decreases with increasing device resistance leaving the device at an elevated temperature relative to the mixing chamber.

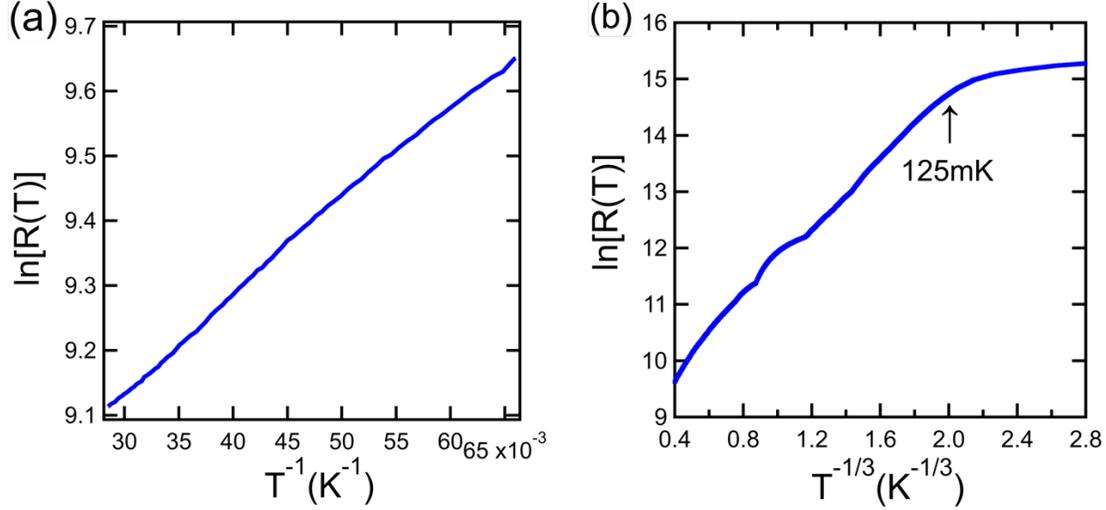

**Figure S4 | Power law T-dependence of CsSnBr$_3$ device outside the VRH-regime.** (a) At high temperatures (35K to 15K), the CsSnBr$_3$ device shows thermally activated transport. (b) At low temperatures (15 K to 125 mK), the device shows 2D Mott VRH transport as described in the main manuscript. Below ~125 mK we find that the device resistance saturates, a phenomenon that is likely associated with thermal decoupling of the device from the cryostat.

**SI 4: Transport measurements on additional epitaxial single-crystal CsSnBr$_3$ device:**

Below we present transport data on an additional device consisting of a 30 nm thick epitaxial layer of CsSnBr$_3$ grown on a single crystal NaCl substrate. The results of these measurements are in good agreement, and reproduce, the results presented for the sample described in the main manuscript.

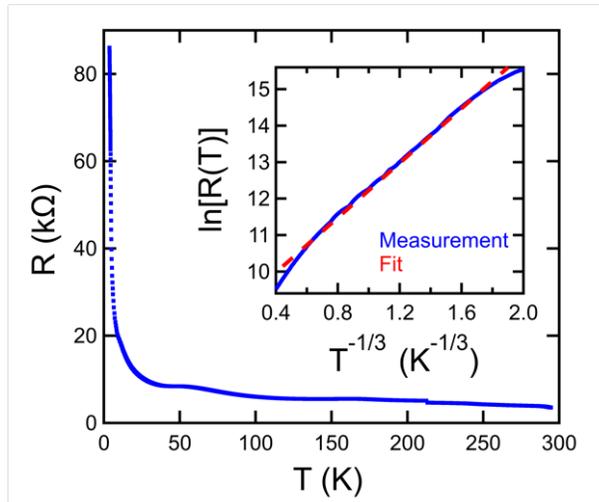

**Figure S5 | Temperature dependent resistance of an additional CsSnBr$_3$ device at B = 0.** The inset shows the logarithm of the resistance as a function of $T^{-1/3}$. The red line is a fit to the data based on the temperature dependence of 2D Mott VRH transport as described in the main manuscript. The characteristic temperature $T_0 = 44K$ is obtained from the slope of the fitted line.

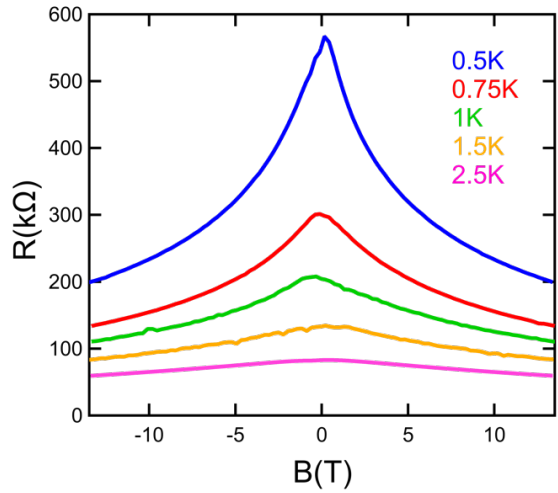

**Figure S6 | Magnetoresistance of an additional CsSnBr$_3$ device.** Giant negative magnetoresistance with no sign of saturation is also observed in another device at various temperatures.

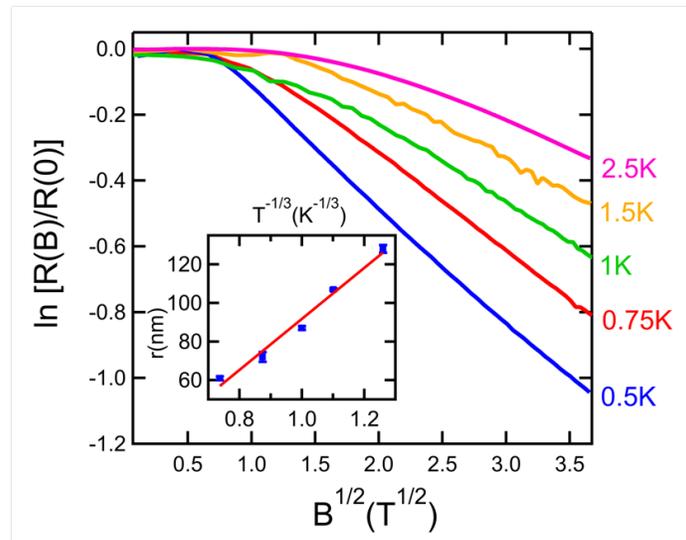

**Figure S7 | Logarithm of the normalized magnetoresistance $\ln[R(B)/R(0)]$ as a function of the square root of the magnetic field $B^{1/2}$ (data is from an additional CsSnBr$_3$ device).** Inset: temperature-dependent hopping length $r(T)$ as a function of $T^{-1/3}$. The corresponding localization length $\xi = 37$ nm in this CsSnBr$_3$ device is derived from the fit shown in the inset (and described in the main manuscript).